\documentclass[conference,10pt]{IEEEtran}
\IEEEoverridecommandlockouts
\usepackage{cite}
\usepackage{amsmath,amssymb,amsfonts}
\usepackage{algorithmic}
\usepackage{graphicx}
\usepackage{textcomp}
\usepackage{xcolor}
\usepackage[hidelinks]{hyperref}
\usepackage{booktabs}
\usepackage{multirow}
\usepackage{enumitem}

\begin{document}

\title{LLMs for Secure Hardware Design and Related Problems:
Opportunities and Challenges}

\author{\IEEEauthorblockN{Johann Knechtel}
\IEEEauthorblockA{New York University Abu Dhabi\\
johann@nyu.edu}
\and
\IEEEauthorblockN{Ozgur Sinanoglu}
\IEEEauthorblockA{New York University Abu Dhabi\\
ozgursin@nyu.edu}
\and
\IEEEauthorblockN{Ramesh Karri}
\IEEEauthorblockA{NYU Tandon School of Engineering\\
rkarri@nyu.edu}}


\maketitle

\begin{abstract}
The integration of Large Language Models (LLMs) into Electronic Design Automation (EDA) and hardware security is rapidly reshaping the semiconductor industry. While LLMs offer unprecedented capabilities in generating
Register Transfer Level (RTL) code, automating testbenches, and bridging the semantic gap between high-level specifications and silicon, they simultaneously introduce severe vulnerabilities. This comprehensive review
provides an in-depth analysis of the state-of-the-art in LLM-driven hardware design, organized around key advancements in EDA synthesis, hardware trust, design for security, and education. We systematically expand on the
methodologies of recent breakthroughs---from reasoning-driven synthesis and multi-agent vulnerability extraction to data contamination and adversarial machine learning (ML) evasion. We integrate general discussions on critical countermeasures, such as dynamic benchmarking to combat data memorization and aggressive red-teaming for robust security assessment. Finally, we synthesize cross-cutting lessons learned to guide future research toward secure, trustworthy, and autonomous design ecosystems.
\end{abstract}

\begin{IEEEkeywords}
Large Language Models, Hardware Security, Electronic Design Automation, Logic Locking, Hardware Trojans, Machine Unlearning, Multi-Agent Systems, Red-Teaming
\end{IEEEkeywords}

\section{Introduction}

The use of generative artificial intelligence (AI), in particular large language models (LLMs), becomes ever-more relevant for electronic design automation (EDA) workflows, showing significant potential to accelerate chip design~\cite{SurveyChipDesign, SurveyEDA}. Given the syntactic similarities between natural language and Hardware Description Languages (HDLs) like Verilog, engineers leverage transformer architectures to automate pattern recognition, code generation, and formal verification~\cite{SurveyChipDesign}. The field has advanced from using LLMs as passive coding assistants to deploying them as active, autonomous agents, using domain-adaptive fine-tuning~\cite{VeriGen} and conversational prompting.

Notwithstanding the progress, the intersection of AI and hardware is a frontier with unique challenges. At the foundation, fine-tuning models on proprietary or public repositories can be  a vector for data-driven vulnerabilities. This includes proprietary Intellectual Property (IP) leaks to end users~\cite{VeriLeaky} and vulnerabilities to backdoor poisoning~\cite{SurveyChipDesign, RTLBreaker}.
Further, the integrity of model evaluation is threatened by data contamination, where models simply memorize static benchmarks rather than learning generalized hardware reasoning~\cite{VeriContaminated}.
Structurally, models face a profound {semantic gap} when attempting to translate high-level constraints into cycle-accurate topologies, impacted by the {representation bottleneck} inherent in compiling natural language into various hardware formats~\cite{RepresentationBottleneck}.
In the security domain, LLMs exhibit an {alignment paradox}, refusing legitimate security analysis while complying with semantically disguised adversarial prompts~\cite{HarmChip}.
Concurrently, traditional learning-based defenses such as Graph Neural Networks (GNNs) are vulnerable to adversarial evasion~\cite{NetDeTox},
and the lowered barrier to design facilitates the rapid generation of stealthy Hardware Trojans (HTs) by malicious actors~\cite{CTF_Hacker}.

Researchers and engineers actively develop a suite of solutions to tackle these challenges.
For example, to bridge the semantic gap and eliminate syntax hallucinations, modern frameworks are transitioning from zero-shot prompting to tool-augmented,
multi-agent loops like AutoChip~\cite{AutoChip}, which dynamically integrate EDA tool feedback (such as compiler diagnostics and simulation logs)  into the reasoning chain. Other frameworks incorporate Monte Carlo
Tree Search (MCTS)~\cite{MCTS_RTL} and formal logic representations like Conjunctive Normal Form (CNF)~\cite{Veritas,RTLforge} to guarantee correctness by construction. To combat data-induced vulnerabilities, machine
unlearning~\cite{SALAD} is a surgical method to sanitize model weights of sensitive IP or triggers. Obsolescence of static datasets is being countered by dynamic, feedback-driven benchmark generation~\cite{TrojanGYM} and adversarial red-teaming~\cite{NetDeTox} to proactively expose and patch detector blind spots.

This paper synthesizes the recent literature to deeply explore these dynamics and map the trajectory of LLM-native and secure EDA flows.
It is organized as follows:
\begin{itemize}[leftmargin=*]
    \item \textit{Section II} details LLM-driven hardware design, spanning reasoning-driven RTL synthesis, testbench generation, and advancing High-Level Synthesis (HLS). It also discusses emerging challenges, including the representation bottleneck and multi-modal configurations.
    \item \textit{Section III} studies threats of LLM-driven design, including backdoors, data contamination, safety misalignments, and IP leakage, and emerging countermeasures.
    \item \textit{Section IV} covers LLM-driven design-for-security, focusing on multi-agent static analysis, red-team assessment,  automating logic locking and side-channel defenses.
    \item \textit{Section V} outlines open-source educational frameworks and Capture-The-Flag (CTF) initiatives.
    \item \textit{Section VI} synthesizes lessons learned to guide future research toward secure, trustworthy ecosystems.
\end{itemize}

\section{LLM-Driven Hardware Design}

Frameworks have evolved from simple coding to tool-augmented generation, verification, and orchestration of multiple models. Selected key advancements are discussed next.

\subsection{From Prompting through Optimized Search to Reasoning}

Early efforts based on simple prompting often resulted in syntactically flawed code. AutoChip~\cite{AutoChip} addressed this by introducing a feedback loop, piping compiler errors back to the agent to autonomously correct
syntax and simulation mismatches. Further optimization was achieved by coupling generation with MCTS~\cite{MCTS_RTL}. This way an  agent may explore and backtrack through a tree of valid Verilog
implementations, optimizing Power, Performance, and Area (PPA) constraints by formulating synthesis as a state-space search. Recent works also use reasoning models, e.g., VeriThoughts~\cite{VeriThoughts} uses DeepSeek-R1-670B to generate Chain-of-Thought traces before emitting Verilog. To combat hallucinations, it uses formal verification for self-consistency checks.

\subsection{Representation: Bottlenecks and Alternatives}

Intermediate representation (IR) can determine end-to-end success more than the LLM choice. A recent study~\cite{RepresentationBottleneck}
evaluates LLMs across six IRs (Verilog, VHDL, Chisel, Bluespec, PyMTL3, and HLS-C) over 202 tasks, revealing large disparities.
Verilog shows simulation pass rates of 83\%--88\%, benefiting from larger training corpora, whereas HLS-C achieves only 3\%--10\% pass rates, suffering from interface protocol mismatches (e.g., start/done handshake) incompatible with standard testbenches. This exposes an {accessibility-competence paradox}: IRs most accessible to ``zero-knowledge'' software engineering (e.g., HLS-C, Python-based PyMTL3) yield the worst quality, due to scarce training data, while HDLs like Verilog and Chisel perform best.

While Verilog remains the dominant IR, RTL++~\cite{RTLpp} addresses the lack of structural awareness in text-only models by encoding RTL into textualized Control Flow Graphs (CFG) and Data Flow Graphs (DFG). Capturing inherent hierarchies and dependencies through these graph representations, the framework outperforms state-of-the-art models on VerilogEval and RTLLM benchmarks.
Taking an alternative approach, Veritas/RTL-Forge~\cite{Veritas,RTLforge} fine-tunes a lightweight model to output CNF clauses, treating the output as a propositional logic mapped via Tseytin transformations. Then, it synthesizes RTL that is correct by construction.

\subsection{Enabling High-Level Synthesis}

There is a gap between software C code and synthesizable HLS-C. C2HLSC~\cite{C2HLSC} tackles this by tasking LLMs to refactor common C code (e.g., removing dynamic memory allocation and pointers)
into HLS-compatible formats. The flow has been proven on C codes for NIST randomness suites and others. For specialized applications, researchers have combined RAG and ReAct pipelines to generate optimized HLS-C/C++ code for neural networks on FPGAs~\cite{NNFPGA}. By feeding Vivado synthesis reports back into context window, LLMs can inject HLS pragmas (e.g., \texttt{PIPELINE}, \texttt{UNROLL}), outperforming automated synthesis tools. Finally, frontier reasoning models can support autonomous HLS Design Space Exploration (DSE) via iterative, constraint-aware pragma insertion~\cite{ReasoningHLS}.

\subsection{Verification: Assertions and Testbenches}

Verification consumes a major part of compute for EDA tooling. On the one hand, automating tasks like assertion generation is invaluable~\cite{SecurityAssertions}; on the other hand,
grounding requirements in RTL semantics remains a key challenge~\cite{viswambharan2026knowledgegraphsmissinglink}.
Studies demonstrate that deterministic generation (e.g., configuring temperature to 0 and top-p to 1) prevents the model from hallucinating constraint scopes, ensuring rigorous SystemVerilog Assertions. To support natural language instructions, researchers proposed a RAG framework~\cite{NL2SVA_RAG} using ``Dynamic Splitting'' to preserve semantic code context and ``HybridRetrieval'' to combine global semantic search with keyword-guided operator retrieval. 

For testbench generation, one can divide complex I/O patterns into manageable sets~\cite{FSM_Testbench}. By feeding coverage metrics (flagging untriggered transitions), the agent expands the testbench to achieve $\sim$100\% state transition coverage on complex Finite State Machines (FSMs).

\subsection{Model Orchestration, Configuration, and Evaluation}

With ever-more models available, orchestration becomes a challenge on its own.
Instead of relying on a single model, VeriDispatcher~\cite{VeriDispatcher} coordinates several backends via pre-inference difficulty prediction. It trains lightweight classifiers on semantic task embeddings to route prompts to the most cost-effective and capable model, yielding better aggregate accuracy while reducing costly API calls by 40\%.
A synthesis-in-the-Loop framework~\cite{HQI_Eval} tested 32 LLMs using the ``Hardware Quality Index'', exposing critical failure modes such as pathological complexity timeouts in frontier LLMs. Other empirical studies~\cite{VeriInteresting} show that, while in-context learning improves structural outputs, domain-specialized fine-tuning is better for complex topologies.

The ``Configuration Over Selection'' study~\cite{ConfigOverSelection} shows that inference-time decoding configuration is vital. Sweeping 108 configurations (vary temperature, top\_p, repetition penalty, presence penalty) across popular open-source models revealed absolute pass-rate gaps of up to 25.5\% points between best and worst settings for the same model. A well-tuned 120B model outperforms a poorly configured 397B model. Moreover, optimal configurations do not transfer: Spearman's rank correlation ($\rho$) of configuration rankings across VerilogEval and RTLLM benchmarks is near zero. These results invalidate the concept of universal default configurations, confirming that open-source LLM performance in EDA requires architecture- and task-aware hyperparameter calibration.

\section{Threats for LLM-Driven Design}

As fabless design houses increasingly rely on automated tools, establishing trust in the models, their training corpora, and their defense mechanisms is critical. Next, we discuss prominent challenges and emerging solutions in some details.

\subsection{Backdoor Attacks}

Relying on public data like GitHub repositories for fine-tuning can expose models to severe data poisoning. The RTL-Breaker framework~\cite{RTLBreaker} demonstrates how malicious actors can systematically inject hidden
triggers into training datasets. Utilizing word-frequency analysis on standard corpora, attackers embed rare keywords (like ``secure'' or ``robust'', pun intended) to serve as activation triggers. When prompted with
these keywords, the backdoored model generates HTs or inferior hardware, like substituting a fast carry-lookahead adder with a slow ripple-carry version. Because the syntax remains valid, these modifications
can evade standard functional checks, highlighting blind spots in current security assessments.

\subsection{Data Contamination}

A pervasive challenge in ML evaluation is data contamination, where benchmark test sets leak into pre-training corpora, artificially inflating capabilities through memorization rather than generalization. The
VeriContaminated framework~\cite{VeriContaminated} utilized established metrics (Min-K\% Prob and Cross-Document Distance metrics) to assess this problem for hardware coding, revealing near-100\% contamination rates in standard benchmarks like VerilogEval across
most commercial models.

To combat this, the community must transition from static test sets to \textit{dynamic benchmarking}, i.e., the continuous, automated generation of private, evolving test sets that models cannot memorize~\cite{LBC,chen-etal-2025-benchmarking-large}. For existing contaminated models, mitigation techniques like threshold-based exclusion of data can filter memorized samples from the training set~\cite{SALAD}. However, applying strict exclusion thresholds inherently degrades overall syntax generation capability, illustrating the tight coupling between memorization and coding fluency.

\subsection{Jailbreaking and Safety (Mis-)Alignment}

Prompt injection in agentic systems is a fundamental architectural vulnerability~\cite{info17010054}.
Current safety alignment mechanisms are trained on general-purpose hazards and fail to comprehend the nuances of hardware security.
The HarmChip framework~\cite{HarmChip} introduces the first domain-specific jailbreak benchmark, spanning 16 hardware security domains (e.g., Cryptography, Advanced Packaging, Firmware) and 120 threats. 
Evaluation across frontier LLMs reveals that keyword-sensitive guardrails indiscriminately refuse legitimate engineering requests (e.g., security audits or defensive analysis) if they contain terms like \textit{Trojan}. Conversely, adversaries can effortlessly bypass these filters via {semantic disguise}, e.g., phrasing a logic-locking attack as PPA-driven resynthesis or Engineering Change Order (ECO). HarmChip exposes significant blind spots, with safety-hardened models exhibiting near-zero resistance to such semantically disguised attacks.

\subsection{Proprietary IP Leakage and Model Sanitization}

Fine-tuning on in-house IP risks leaking sensitive architectural blueprints to end-users. VeriLeaky~\cite{VeriLeaky} quantified this threat on a large-scale case study, revealing that fine-tuned models can flawlessly
regenerate industrial-grade cryptographic accelerators and other sensitive IP.
The study explored prompt variations, proving that revealing minor structural hints like interface declarations can significantly exacerbate formal-equivalence leakage of IP. While pre-applying logic locking to the
training data reduces this leakage, it also degrades the model's general utility, calling for more sophisticated defenses.

To resolve this without retraining, the SALAD framework~\cite{SALAD} implements machine unlearning. By evaluating established gradient-based and preference-based algorithms on a range of fine-tuned models, SALAD determined that methods like SimNPO can surgically scrub sensitive IP, contaminated benchmarks, or malicious backdoors while preserving robust Verilog coding performance.

\section{LLM-Driven Design for Security}

Among other trends, we note that agents are becoming sophisticated orchestrators capable of implementing and auditing (hardware) security problems for various domains. Next, we discuss related challenges and solutions in details.

\subsection{IP Protection: Logic Locking and Redaction}

Logic locking protects design IP via key-dependent gates. GLLaMoR~\cite{GLLaMoR} converts netlists into adjacency lists and prompts models to perform topological reasoning (Depth-First Search) to identify crucial nodes,
achieving significant speedups over traditional algorithmic analysis. Hector~\cite{Hector} advances this effort through an agentic feedback loop, including equivalence checks for functional correctness and SAT
attacks for resilience.
LockForge~\cite{LockForge} solves the difficult ``paper-to-code'' challenge with a multi-agent framework (Coder, Judge, Examiner). It parses academic PDFs, implements complex locking schemes in Python, and validates them via strict similarity scoring.
Similarly, \cite{ghimire2026agentssecurehardwareevaluating} introduces a framework that combines retrieval-grounded planning with structured lock-plan generation.
The agentic approach utilizes SAT-based security evaluation to iteratively refine implementations.

For embedded FPGAs used for IP protection via redaction, ARIANNA~\cite{ARIANNA} automates DSE.
It clusters candidate redaction modules and uses a branch-and-bound algorithm to determine optimal eFPGA configurations (K and N logic block parameters). This fine-tuning reduces area overheads by up to 3.3$\times$ while maximizing fabric use.

\subsection{Side-Channel Mitigation and Post-Quantum Cryptography}

Pre-silicon Side-Channel Analysis (SCA) traditionally requires exhaustive power simulations.
NetlistWhisperer~\cite{NetlistWhisperer} utilizes an ensemble of fine-tuned GPT models to analyze gate-level netlists and predict Test Vector Leakage Assessment (TVLA) bounds. By employing ensemble voting to overcome the severe class imbalance of rare leaky nets, it accurately classifies vulnerabilities and generates secured HDL implementing Domain-Oriented Masking. 

For Post-Quantum Cryptography (PQC), LLM4PQC~\cite{LLM4PQC} uses an agentic workflow.
Prompt constraints, formulated as acting as a Lead Cryptographic Engineer, force the agent to extract critical kernels, enforce constant-time execution, and apply HLS static array mapping to bypass dynamic memory limitations, among other steps.
This workflow can yield area-efficient implementations. In an extension,
the flow offers SCA-resilient accelerators~\cite{vts26_special_session_secure_PQC}.

\subsection{Trojan Detection}

Traditional HT detectors often utilize GNNs operating on DFGs or similar concepts. However, graph conversion inherently discards rich textual semantics, variable intent, and control flow. TrojanLoC~\cite{TrojanLoC} bypasses graph conversion entirely, processing raw Verilog. Utilizing the expansive TrojanInS dataset, it extracts module-level and fine-grained line-level embeddings directly from the code using an RTL-finetuned transformer. Classified via lightweight gradient-boosting trees, TrojanLoC achieves a 99\% F1-score for module detection and unprecedented line-level localization of malicious payloads, proving structural graph constraints are unnecessary when leveraging semantic comprehension.

\subsection{Red-Teaming and Adversarial Evasion}

Security assessment in EDA tooling increasingly relies on ML to detect piracy or Trojans.
However, evaluating the robustness of these defenses necessitates rigorous \textit{red-teaming}, i.e., systematically challenging systems with adversarial examples, to expose algorithmic
blind spots or vulnerabilities.

NetDeTox~\cite{NetDeTox} advances red-teaming efforts to evade GNN piracy detectors. 
Instead of brute-force global rewiring, NetDeTox orchestrates a hybrid Reinforcement Learning (RL) and LLM pipeline. The RL agent groups gates into feature-based buckets and then samples high-leverage subnetlists. The
LLM acts as a contextual planner, strictly following a decision sequence to generate localized rewriting plans. By applying targeted gate substitutions, NetDeTox bypasses GNN-based detectors across 90\% of test cases.
Strikingly, this localized red-teaming often achieves \textit{negative} area overhead, producing adversarial circuits that are fully outperforming prior state-of-the-art~\cite{AttackGNN} and even the baseline circuits.
Similarly, TrojanGYM~\cite{TrojanGYM} orchestrates an attack-defense loop to evade HT detection.
An agent generates HTs; these designs are functionally verified and evaluated by a robust GNN detector. The continuous detection scores serve as feedback, forcing the agent to iteratively restructure the Abstract Syntax Tree (AST) until the Trojan evades detection, achieving 83\% evasion rates and systematically mapping detector blind spots.

\subsection{Bug Detection and Code Analysis}

Detecting subtle RTL bugs requires deep contextual understanding. Common Weakness Enumerations (CWEs) present a unique challenge because hardware vulnerabilities often overlap hierarchically and manifest differently depending on micro-architectural context, rendering standard pattern-matching brittle. 
To overcome this, {VeriCWEty}~\cite{VeriCWEty} pioneers an embedding-enabled detection framework. Rather than explicitly extracting ASTs or relying on flat classification, VeriCWEty puts HDL through a Verilog-finetuned decoder. It extracts dense vector embeddings at both the global module level and the line level. By utilizing a majority-voting ensemble of frontier models to curate high-quality training labels, a simple XGBoost classifier trained on these embeddings achieves 89\% precision in identifying critical vulnerabilities (like CWE-1244 and CWE-1245) and 96\% accuracy in pinpointing the line-level location of the bug~\cite{VeriCWEty}. 

LASHED~\cite{LASHED} combines generative AI with static analysis. The deterministic tool flags violations, and AI contextualizes them against CWEs to filter false positives. MARVEL~\cite{MARVEL} employs a hierarchical Multi-Agent architecture where a Supervisor delegates tasks to  executors (Linter, CWE, and Similar-Bug RAG agents). Finally, FLAG~\cite{FLAG} introduces test-free fault localization, identifying anomalous line-level logic directly from source code by calculating token-level generation probabilities (logprob) and embedding distances.

\begin{figure*}[tb]
\hspace{-5mm}
    \includegraphics[trim={15mm 15mm 15mm 55mm},clip,width=1.05\textwidth]{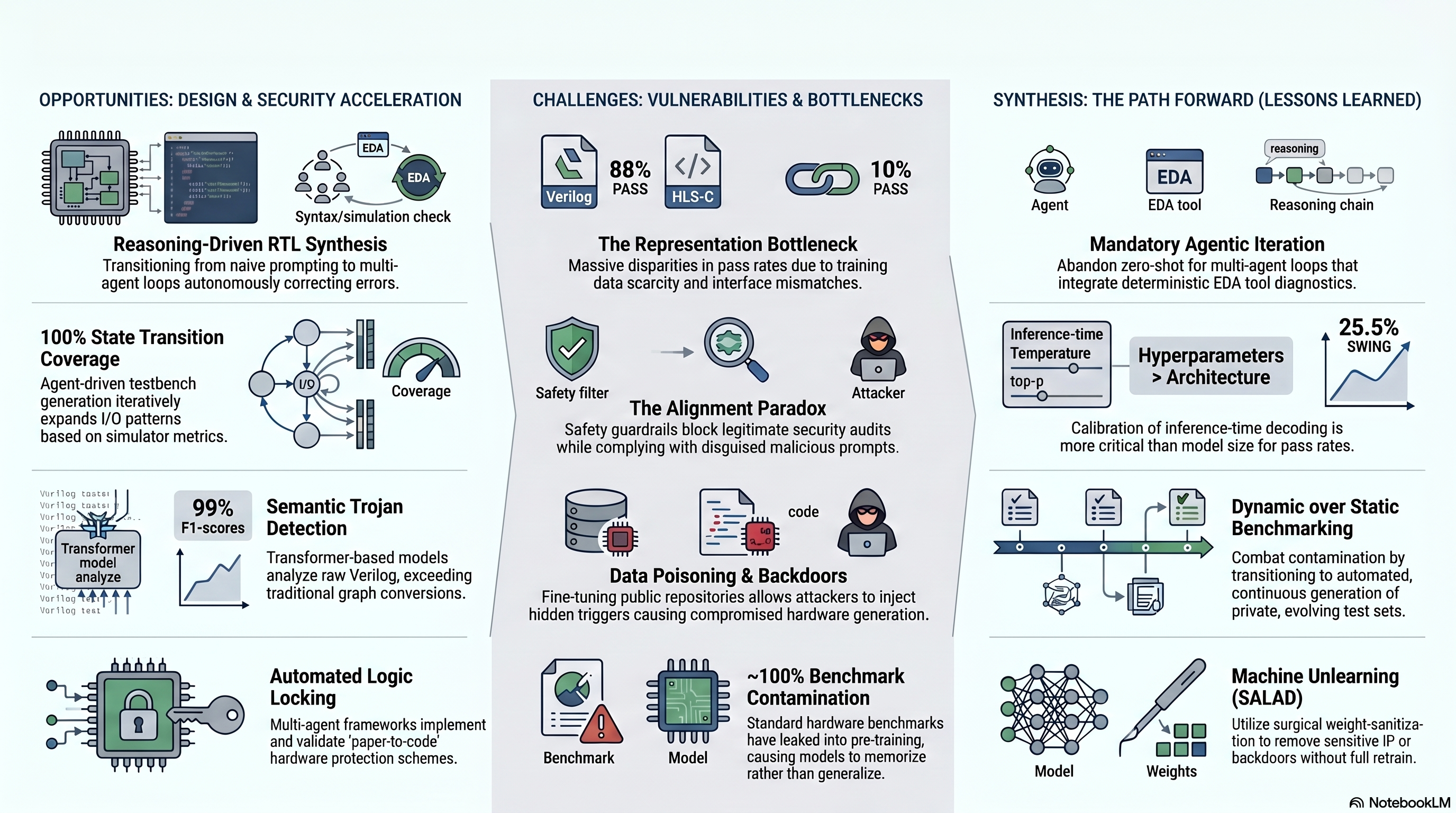}
    \caption{Selected opportunities and challenges as well as lessons learned for LLM-driven and trustworthy hardware design.
	    Image generated using NotebookLM.}
    \label{fig:lessons}
\end{figure*}

\section{Education and Outreach}

The rapid integration of generative AI into EDA requires academic curricula and learning environments to radically adapt. Next, we discuss latest efforts towards that end.

\subsection{Modular Courseware: GUIDE}

To standardize education, the GUIDE framework~\cite{GUIDE} provides an open-source repository translating state-of-the-art research into runnable Google Colab environments. Organized into standardized units (slides, videos, self-contained code), GUIDE enables instructors to construct tailored courses. By mandating standardized deliverables (e.g., waveforms, EDA tool logs), it enforces rigorous engineering practices alongside automated workflows, allowing students to securely practice HT insertion and logic locking evasion.

\subsection{Scalable CTFs and Evaluation}

Capture-the-Flag  competitions are crucial testbeds for security education. A cross-regional study on the CSAW CTF formulated scalable principles for AI-assisted environments~\cite{AgenticCTFEdu}.

To rigorously evaluate these competitions, organizers formalized autonomy levels---Human-in-the-Loop, Autonomous Agent, and Hybrid---and introduced the CTFJudge framework~\cite{CTF_Judge} alongside the CTFTiny benchmark.  This infrastructure computes a ``CTF Competency Index'' by mandating traceable evidence (conversational logs and agent execution trajectories) to verify the reasoning-action-output chain.
Analysis revealed that, while autonomous workflows outperform humans on complex tool interactions, beginners strongly preferred simple single-agent loops augmented with strict prompt checklists, providing a roadmap for lowering the barrier to entry in secure hardware education.

Analyzing how humans collaborate with LLM tools to exploit systems, the ``Lowering the Bar'' study~\cite{CTF_Hacker} evaluated a global AI Hardware Attack Challenge. Results demonstrated that competitors with minimal hardware expertise successfully used prompt-engineered agents to locate vulnerabilities and insert severe Trojans (e.g., side-channel leaks and denial-of-service payloads), underscoring  robust alignment guardrails.

\section{Lessons Learned and Conclusion}

Several cross-cutting paradigms defining the future of secure design automation are listed next and illustrated in Fig.~\ref{fig:lessons}.

\begin{itemize}[leftmargin=*]
    \item \textit{Agentic Iteration:} Zero-shot prompting fails for complex hardware topologies. State-of-the-art frameworks require iterative, multi-agent loops tightly integrated with deterministic EDA tools.
    \item \textit{Hyperparameters vs Architectures:} Model capabilities can be largely masked by default configurations. Hyperparameter tuning can induce significant pass-rate swings,
making configuration as vital as model selection.
    \item \textit{Data Modality vs Capability:} Models tuned on Verilog are superior for semantic understanding and fine-grained tasks, while adjacency lists and graphs can excel for structural and topological analysis.
    \item \textit{Representation Bottleneck:} The target IR fundamentally bounds generation quality. For example, HLS-C often fails due to interface mismatches, while Verilog offers sufficient data richness for training.
Advanced coding requires navigating the accessibility-competence paradox.
    \item \textit{Alignment Paradox:} General-purpose safety filters fail in hardware contexts. They over-refuse legitimate security audits while remaining entirely vulnerable to semantically disguised adversarial engineering requests.
    \item \textit{Red-Teaming:} ML-based security defenses can project a false sense of security unless subjected to rigorous testing. Red-teaming via LLM-driven frameworks ensure defenses are robust against real-world attackers.
    \item \textit{Dynamic vs Static Benchmarking:} Due to pervasive data contamination and rapid adaptive evasion capabilities, static datasets must be replaced by feedback-driven, dynamically generated benchmarks.
\end{itemize}
Overall, while LLMs provide extraordinary acceleration for EDA workflows, they necessitate a fundamentally new paradigm of hardware security relying on orchestrated agents operating within rigorously defined and verified boundaries.

\IEEEtriggeratref{30}
\bibliographystyle{IEEEtran}
\bibliography{references}

\end{document}